# GOVERNANCE OF THE INTERNET OF THINGS (IoT)
[Pre-Publication Draft]

Lawrence J. Trautman*
Mohammed T. Hussein**
Louis Ngamassi***
Mason J. Molesky****

\* BA, The American University; MBA, The George Washington University; J.D., Oklahoma City University School of Law. Mr. Trautman is Associate Professor of Business Law and Ethics at Prairie View A&M University. He is a past president of the New York and Washington/Baltimore Chapters of the National Association of Corporate Directors (NACD). He may be contacted at Lawrence.J.Trautman@gmail.com.

\*\* BS, Prairie View A&M University; M.S. Texas A&M University-Kingsville (electrical engineering); Ph. D., Texas A&M University (electrical engineering). Dr. Hussein is Assistant Professor of Management Information Systems (MIS) at Prairie View A&M University. He may be contacted at mthussein@pvamu.edu.

\*\*\* BS, University of Yaoundé; MS, Pace University- New York (Computer Information Systems); Ph.D. Penn State University (Information Sciences and Technology). Dr. Ngamassi is an Associate Professor of Management Information Systems (MIS) at Prairie View A&M University and a Faculty Affiliate of the Hazard Reduction and Recovery Center at Texas A&M University, College Station. He may be contacted at longamassi@pvamu.edu.

\*\*\*\* BS, Alma College (mathematics and computer science); M.S. (cybersecurity), Ph.D. Candidate (computer science), The George Washington University. Mr. Molesky is an adjunct professor at The George Washington University. He may be contacted at masonmolesky@gmail.com.

ABSTRACT

Today's increasing rate of technological change results from the rapid growth in computer processing speed, when combined with the cost decline of processing capacity, and is of historical import. The daily life of billions of individuals worldwide has been forever changed by technology in just the last few years. Costly data breaches continue at an alarming rate. The challenge facing humans as they attempt to govern the process of artificial intelligence, machine learning, and the impact of billions of sensory devices connected to the Internet is the subject of this Article.

We proceed in nine sections. First, we define the Internet of Things (IoT), comment on the explosive growth in sensory devices connected to the Internet, provide examples of IoT devices, and speak to the promise of the IoT. Second, we discuss legal requirements for corporate governance as a foundation for considering the challenge of governing the IoT. Third, we look at potential IoT threats. Fourth, we discuss the Mirai botnet. Fifth, is a look at IoT threat vector vulnerabilities during times of crisis. Sixth, we discuss the Manufactured Usage Description (MUD) methodology. Seventh, is a discussion of recent regulatory developments. Next, we look at a few recommendations. And finally, we conclude. We believe this Article contributes to our understanding of the widespread exposure to malware associated with IoT and adds to the nascent but emerging literature on governance of enterprise risk, a subject of vital societal importance.

CONTENTS

GOVERNANCE OF THE INTERNET OF THINGS (IoT)

> *We're entering an age of acceleration. The models underlying society at every level, which are largely based on a linear model of change, are going to have to be redefined. Because of the explosive power of exponential growth, the twenty-first century will be equivalent to 20,000 years of progress at today's rate of progress; organizations have to be able to redefine themselves at a faster and faster pace.*
>
> *Ray Kurzweil*
> *Director of Engineering at Google[1]*

I.     OVERVIEW

Today's increasing rate of technological change results from the rapid growth in computer processing speed, when combined with the cost decline of processing capacity, and is of historical import.[2] Giaretta, Dragoni and Massacci report, "Smart homes are equipped with a growing number of IoT devices that capture more and more information about human beings lives. However, manufacturers paid little or no attention to security…"[3] As the U.S. National Institutes of Standards and Technology reports in their July 2019 exposure draft titled *Core Cybersecurity feature Baseline for Securable IoT Devices: A Starting Point for IoT Device Manufacturers*:

> Manufacturers are creating an incredible variety and volume of Internet of Things (IoT) devices, which incorporate at least one transducer (sensor or actuator) for interacting directly with the physical world, have at least one network interface (e.g., Ethernet, WiFi, Bluetooth, Long-Term Evolution [LTE], ZigBee), and are not conventional IT devices for which the identification and implementation of cybersecurity features is already well understood (e.g., smartphone, laptop). Many IoT devices provide computing functionality, data storage, and network connectivity for equipment that previously lacked these functions. In turn, these functions enable new efficiencies and technological capabilities for the equipment, such as remote access for monitoring, configuration, and troubleshooting. IoT can also add the ability to analyze data about the physical world and use the results to better inform decision making, alter the physical environment, and anticipate future events.[4]

---

[1] Lawrence J. Trautman, *Bitcoin, Virtual Currencies and the Struggle of Law and Regulation to Keep Pace,* 102 MARQ. L. REV. 447, 470 (2018), https://ssrn.com/abstract=3182867, *citing* THOMAS L. FRIEDMAN, THANK YOU FOR BEING LATE: AN OPTIMIST'S GUIDE TO THRIVING IN THE AGE OF ACCELERATIONS 187 (2016).
[2] *Id.*
[3] *See* Alberto Giaretta, Nicola Dragoni & Fabio Massacci, *Protecting the Internet of Things with Security-by-Contract and Fog Computing,* IEEE Consumer Electronics Magazine, 413arXiv:1903.04794v2 [cs.CR] (May 2, 2019).

The daily life of billions of individuals worldwide has been forever changed by technology in just the last few years.[5] Costly data breaches continue at an alarming rate.[6] By 2020, "IoT devices are increasingly being implicated in cyber-attacks, raising community concern about the risks they pose to critical infrastructure, corporations, and citizens."[7] In an effort to mitigate "this risk, the IETF is pushing IoT vendors to develop formal specifications of the intended purpose of their IoT devices, in the form of a Manufacturer Usage Description (MUD), so that their network behavior in any operating environment can be locked down and verified rigorously."[8] The challenge facing humans as they attempt to govern the process of artificial intelligence, machine learning, and the impact of billions of sensory devices connected to the Internet is the subject of this Article.

---

[4] *See* Michael Fagan, Katerina N. Megas, Karen Scarfone & Matthew Smith, *Core Cybersecurity feature Baseline for Securable IoT Devices: A Starting Point for IoT Device Manufacturers,* Draft NISTIR 8259 (July 2019), https://nvlpubs.nist.gov/nistpubs/ir/2019/NIST.IR.8259-draft.pdf.

[5] Some examples of these recent technological advances having profound impact include: Google (founded 1998), *see* Lawrence J. Trautman, *How Google Perceives Customer Privacy, Cyber, E-commerce, Political and Regulatory Compliance Risks*, 10 WM. & MARY BUS. L. REV.1 (2018), https://ssrn.com/abstract=3067298 [https://perma.cc/23UM-L4Z4] (citing Alphabet Inc., Quarterly Report, (Form 10-Q), at 17 (Oct. 27, 2017), PayPal, Registration Statement (Form S-1), at 9 (June 12, 2002), https://www.sec.gov/Archives/edgar/data/1103415/000091205702023923/a2082068zs-1.htm [https://perma.cc/7G9F-E8FL]; Lawrence J. Trautman, *E-Commerce, Cyber, and Electronic Payment System Risks: Lessons from PayPal*, 17 U.C. DAVIS BUS. L.J. 261, 274–79 (2016); Facebook (2004), bitcoin, blockchain and virtual currencies (2009), Lawrence J. Trautman, *Is Disruptive Blockchain Technology the Future of Financial Services?,* 69 CONSUMER FIN. L.Q. REP. 232, 234 (2016), http://ssrn.com/abstract=2786186; *See* Lawrence J. Trautman & Alvin Harrell, *Bitcoin Versus Regulated Payment Systems: What Gives?* 38 CARDOZO L. REV. 1041, 1055 (2017), http://ssrn.com/abstract=2730983; Lawrence J. Trautman, *Virtual Currencies: Bitcoin & What Now After Liberty Reserve, Silk Road, and Mt. Gox?,* 20 RICHMOND J. L. & TECH. 13, 43 (2014), http://ssrn.com/abstract=2393537; Uber (2009), *see Company Info*, UBER, https://www.uber.com/newsroom/company-info/ [https://perma.cc/UDE2-9L54] (last visited Jan. 19, 2018); WhatsApp (2009), Parmy Olson, *Exclusive: The Rags-to-Riches Tale of how Jan Koum Built WhatsApp Into Facebook's New $19 Billion Baby*, FORBES (Feb. 19, 2014, 7:58 PM), https://www.forbes.com/sites/parmyolson/2014/02/19/exclusive-inside-story-how-jan-koum-built-whatsapp-into-facebooks-new-19-billion-baby/ [https://perma.cc/ZRM5-22WM]; Instagram (2010), *see* Gwyn Topham, *Look Ma, Mo Hands: What will it Mean when all Cars can Drive Themselves?*, GUARDIAN (United Kingdom) (Nov. 25, 2017), https://www.theguardian.com/business/2017/nov/25/autonomous-vehicles-when-all-cars-drive-themselves-what-will-it-mean [https://perma.cc/8XRJ-3KNT].

[6] *See* BRUCE MIDDLETON, A HISTORY OF CYBER SECURITY ATTACKS: 1980 TO PRESENT (Taylor & Francis, 2017); § IV *Infra*.

We proceed in nine sections. First, we define the Internet of Things (IoT), comment on the explosive growth in sensory devices connected to the Internet, provide examples of IoT devices, and speak to the promise of the IoT. Second, we discuss legal requirements for corporate governance as a foundation for considering the challenge of governing the IoT. Third, we look at potential IoT threats. Fourth, we discuss the Mirai botnet. Fifth, is a look at IoT threat vector vulnerabilities during times of crisis. Sixth, we discuss the Manufactured Usage Description (MUD) methodology. Seventh, is a discussion of recent regulatory developments. Next, we look at a few recommendations. And finally, we conclude. We believe this Article contributes to our understanding of the widespread exposure to malware associated with IoT and adds to the nascent but emerging literature on governance of enterprise risk, a subject of vital societal importance.

---

[7] *See* Ayyoob Hamza, Dinesha Ranathunga, Hassan Habibi Gharakheili, Matthew Roughan & Vijay Sivaraman, *Clear as MUD: Generating, Validating and Applying IoT Behavioral Profiles,* IoT S&P'18, Aug. 20, 2018, Budapest, Hungary. *See also* Sarah Coble, *Amazon Doorbell Camera Lets Hackers Access Household Network,* INFOSECURITY MAGAZINE (Nov. 7, 2019); Angella Foster, *When Parents Spy on Nannies,* Op-Ed, N.Y. TIMES, Aug. 20, 2019 at A23; Sandra E. Garcia, *Maker of Popular Home Security Camera Exposes Users,* N.Y. TIMES, Dec. 20, 2019 at B4; Kate Murphy, *Is There a Tiny Spy In Your TV Room? How to Unmask It,* N.Y. TIMES, Nov. 16, 2019 atB1; Zack Whittaker, *Amazon Ring Doorbells Expose Home Wi-Fi Passwords to Hackers,* TECHCRUNCH (Nov. 7, 2019).

[8] *See* Hamza, *supra* note 7.

## II. THE INTERNET OF THINGS (IoT)

*"Continued rapid technological progress remains central to economic prosperity and social well-being, but it is also introducing potential new threats. The Internet of Things (IoT) is connecting billions of new devices to the Internet, but it also broadens the attack potential of cyber actors against networks and information."*

*Daniel R. Coats*
*Director of National Intelligence*
*May 11, 2017[9]*

The Internet of Things (IoT) can be defined as "a vast network of devices that are connected to the internet and, consequently, each other increasingly."[10] In the simplest of terms, any sensory device that may find connectivity to the Internet is a part of the Internet of Things. Included are wearables such as: watches or any sensory devices that are worn on clothing to detect health vitals such as heart rate, body temperature, blood pressure; smart phones; all those household devices such as front door video cameras, plant and gardening watering needs, voice command devices to control televisions, room temperature, etc. Military sensory applications are robust: IoT remote sensory devices monitor troop and vehicle movement; sonar and space sensory applications; vital signs of healthy and wounded troops in the battle theatre, just to name a few.

---

[9] Worldwide Threat Assessment of the U.S. Intelligence Community, Before the S. Select Comm. on Intelligence (115th Cong. 3 (2017) (statement of Daniel R. Coats, Director of National Intelligence).

[10] NIST, *Securing Small Business and Home Internet of Things (IoT) Devices: Mitigating Network-Based Attacks By Using Manufactured Usage Description,* Nat'l Cybersecurity Center of Excellence (Feb. 2019),

Bruce Sinclair writes that, "the Internet of Things (IoT) is just an evolution of the Internet. No more, no less. But the business ramifications of IoT are revolutionary and will usher in the Outcome Economy."[11] Mr. Sinclair further observes that "The Internet of Things killer app is outcomes. It's outcomes that customers usually want. They don't even care about products: they care about what products do for them…. Customers don't want to own cars, they want to get from one place to another, fast and safe."[12]

Trautman and Ormerod write that "The proliferation of novel consumer devices and increased Internet-dependent business and government data systems introduces vulnerabilities of unprecedented magnitude."[13] These "Digital vulnerabilities touch upon a number of different areas of the law: privacy,[14] risk management,[15] corporate governance[16] (including the duties of care,[17] monitor,[18] and disclosure[19]), breach notification,[20] information and data security,[21] securities regulation,[22] law of war,[23] constitutional provisions,[24] and more.[25]

---

[11] BRUCE SINCLAIR, IOT INC.: HOW YOUR COMPANY CAN USE THE INTERNET OF THINGS TO WIN IN THE OUTCOME ECONOMY xi (McGraw Hill Education, 2017).
[12] *Id.* at xxii.
[13] *See* Lawrence J. Trautman & Peter C. Ormerod, *Industrial Cyber Vulnerabilities: Lessons from Stuxnet and the Internet of Things,* 72 U. MIAMI L. REV. 761, 764 (2018), http://ssrn.com/abstract=2982629; *citing* Trey Herr & Allan A. Friedman, Redefining Cybersecurity, American Foreign Policy Council - Defense Technology Program Brief, No. 8, (2015), https://ssrn.com/abstract=2558265; Daniel J. Solove, Identity Theft, Privacy, and the Architecture of Vulnerability, 54 HASTINGS L.J. 1227 (2003), https://ssrn.com/abstract=416740.
[14] *See* Trautman & Ormerod, *supra* note 13, *citing* Corey Ciocchetti, The Privacy Matrix, 12 U. FLA. J. TECH. L. & POL'Y, __ (2008), https://ssrn.com/abstract=1090423; Neil M. Richards & Jonathan H. King, Big Data and the Future for Privacy, In Handbook of Research on Digital Transformations (Elgar 2016), https://ssrn.com/abstract=2512069; Sasha Romanosky & Alessandro Acquisti, *Privacy Costs and Personal Data Protection: Economic and Legal Perspectives,* 24 BERKELEY TECH. L.J. __ (2009), https://ssrn.com/abstract=1522605; Daniel J. Solove & Woodrow Hartzog, *The FTC and the New Common Law of Privacy,* 114 COLUMBIA L. REV. 583 (2014), https://ssrn.com/abstract=2312913; Daniel J. Solove & Chris Jay Hoofnagle, A Model Regime of Privacy Protection (Version 2.0), GWU Law School Public Law Research Paper No. 132; GWU Legal Studies Research Paper No. 132, https://ssrn.com/abstract=699701; Robert Kirk Walker, *The Right to be Forgotten,* 64 HASTINGS L. J. 257 (2012), https://ssrn.com/abstract=2017967
[15] Trautman & Ormerod, *supra* note 13, *citing* Liam M. D. Bailey, *Mitigating Moral Hazard in Cyber-Risk Insurance,* 3 J.L. & CYBER WARFARE 1 (2014), https://ssrn.com/abstract=2424958; Shauhin A. Talesh, *Data Breach, Privacy, and Cyber Insurance*, __ LAW & SOCIAL INQUIRY __ (2017), https://ssrn.com/abstract=2974233; Lawrence J. Trautman & Kara Altenbaumer-Price, *D&O Insurance: A Primer,* AM. U. BUS. L. REV. 337 (2012), http://ssrn.com/abstract=1998080.

---

[16] Trautman & Ormerod, *supra* note 13, *citing* John Armour, Henry Hansmann & Reinier Kraakman, *Agency Problems, Legal Strategies, and Enforcement,* Oxford Legal Studies Research Paper No. 21/2009; Yale Law, Economics & Public Policy Research Paper No. 388; Harvard Law and Economics Research Paper Series No. 644 ; ECGI - Law Working Paper No. 135/2009, https://ssrn.com/abstract=1436555; Lucian A. Bebchuk, Alma Cohen & Allen Ferrell, *What Matters in Corporate Governance?,* 22 REV. FIN. STUD. 783 (2009), https://ssrn.com/abstract=593423; Lawrence J. Trautman & Kara Altenbaumer-Price, *The Board's Responsibility for Information Technology Governance*, 28 J. MARSHALL J. COMPUTER & INFO. L. 313 (2011), http://www.ssrn.com/abstract=1947283.

[17] Trautman & Ormerod, *supra* note 13, *citing* Stephen M. Bainbridge, Star Lopez & Benjamin Oklan, The Convergence of Good Faith and Oversight, UCLA School of Law, Law-Econ Research Paper No. 07-09, https://ssrn.com/abstract=1006097; Melvin A. Eisenberg, The Duty of Good Faith in Corporate Law, 31 DEL. J. CORP. L. 1 (2005), https://ssrn.com/abstract=899212.

[18] *Id., citing* Robert T. Miller, *The Board's Duty to Monitor Risk after Citigroup,* 12 U. PA. J. BUS. L. 1153 (2010), https://ssrn.com/abstract=1696166.

[19] Trautman & Ormerod, *supra* note 13, *citing* Bernard S. Black, *The Core Fiduciary Duties of Outside Directors*, __ ASIA BUS. L. REV. 3 (2001), https://ssrn.com/abstract=270749; Henry T. C. Hu, *Too Complex to Depict? Innovation, 'Pure Information,' and the SEC Disclosure Paradigm*, 90 TEX. L. REV. __ (2012), https://ssrn.com/abstract=2083708; Peter A. Swire, *Theory of Disclosure for Security and Competitive Reasons: Open Source, Proprietary Software, and Government Agencies,* 42 HOU. L. REV. 101 (2006), https://ssrn.com/abstract=842228.

[20] Trautman & Ormerod, *supra* note 13, *citing* Fabio Bisogni, *Evaluating Data Breach Notification Laws - What Do the Numbers Tell Us?,* TPRC 41: The 41st Research Conference on Communication, Information and Internet Policy, *available at* https://ssrn.com/abstract=2236144; Dana Lesemann, *Once More Unto the Breach: An Analysis of Legal, Technological and Policy Issues Involving Data Breach Notification Statutes,* 4 AKRON INTELL. PROP. J. 203 (2010), https://ssrn.com/abstract=1671082; Paul M. Schwartz & Edward J. Janger, Notification of Data Security Breaches. 105 MICH. L. REV. 913 (2007), https://ssrn.com/abstract=908709; Jane K. Winn, Are 'Better' Security Breach Notification Laws Possible?, 24 BERK. TECH. L. J. __ (2009), https://ssrn.com/abstract=1416222.

[21] Trautman & Ormerod, *supra* note 13, *citing* Ian Brown, Lilian Edwards & Christopher Marsden, Information Security and Cybercrime, In LAW AND THE INTERNET, (3rd Ed., L. Edwards, C. Waelde, eds., Oxford: Hart, 2009), https://ssrn.com/abstract=1427776; Paul Ohm, Broken Promises of Privacy: Responding to the Surprising Failure of Anonymization , 57 UCLA L. REV. 1701 (2010), https://ssrn.com/abstract=1450006; Daniel J. Solove, The New Vulnerability: Data Security and Personal Information, In SECURING PRIVACY IN THE INTERNET AGE, (Radin & Chander, eds., Stanford University Press, 2008), https://ssrn.com/abstract=583483; Richard Warner & Robert H. Sloan, Defending Our Data: The Need for Information We Do Not Have, (2016), https://ssrn.com/abstract=2816010; Josephine Wolff, Models for Cybersecurity Incident Information Sharing and Reporting Policies, TPRC 43: The 43rd Research Conference on Communication, Information and Internet Policy Paper (2014), https://ssrn.com/abstract=2587398.

[22] Trautman & Ormerod, *supra* note 13, *citing* Zohar Goshen & Gideon Parchomovsky, *The Essential Role of Securities Regulation*, 55 DUKE L.J. 711 (2006), https://ssrn.com/abstract=600709; Andrea M. Matwyshyn, *Material Vulnerabilities: Data Privacy, Corporate Information Security and Securities Regulation,* 3 BERKELEY BUS. L.J. 129 (2005), https://ssrn.com/abstract=903263; Robert B. Thompson & Hillary A. Sale, *Securities Fraud as*

Explosive Growth in Sensory Devices

The use of "IoT products at home and work… [anticipates] the number of connected devices to reach 20.4 billion by 2020."[26] The National Institutes of Standards and Technology (NIST) observes:

> The Internet of Things (IoT) is a rapidly evolving and expanding collection of diverse technologies that interact with the physical world. IoT devices are an outcome of combining the worlds of information technology (IT) and operational technology (OT). Many IoT devices are the result of the convergence of cloud computing, mobile computing, embedded systems, big data, low-price hardware, and other technological advances. IoT devices can provide computing functionality, data storage, and network connectivity for equipment that previously lacked them, enabling new efficiencies and technological capabilities for the equipment, such as remote access for monitoring, configuration, and troubleshooting. IoT can also add the abilities to analyze data about the physical world and

---

*Corporate Governance: Reflections Upon Federalism,* VANDERBILT L. REV. (2003), https://ssrn.com/abstract=362860; Lawrence J. Trautman & George P. Michaely, *The SEC & The Internet: Regulating the Web of Deceit,* 68 CONSUMER FIN. L.Q. REP. 262 (2014), http://www.ssrn.com/abstract=1951148.

[23] *See* Steven M. Bellovin, Susan Landau & Herbert S. Lin, Limiting the Undesired Impact of Cyber Weapons: Technical Requirements and Policy Implications, https://ssrn.com/abstract=2809463; Eric Talbot Jensen, The Tallinn Manual 2.0: Highlights and Insights, __ GEORGETOWN J. INT'L L. (Forthcoming), https://ssrn.com/abstract=2932110; Eric Talbot Jensen, Computer Attacks on Critical National Infrastructure: A Use of Force Invoking the Right to Self-Defense, 38 STANFORD J. INT'L L. 207 (2002), https://ssrn.com/abstract=987046; Dakota S. Rudesill, James Caverlee & Daniel Sui, *The Deep Web and the Darknet: A Look Inside the Internet's Massive Black Box, Woodrow Wilson International Center for Scholars,* STIP 03, October 2015; Ohio State Public Law Working Paper No. 314, https://ssrn.com/abstract=2676615; Michael N. Schmitt & Sean Watts, The Decline of International Humanitarian Law Opinio Juris and the Law of Cyber Warfare, 50 TEX. INT'L L. J. __ (2015), https://ssrn.com/abstract=2481629; Scott Shackelford, Scott Russell & Andreas Kuehn, Unpacking the International Law on Cybersecurity Due Diligence: Lessons from the Public and Private Sectors, __CHI. J. INT'L L. (2016), https://ssrn.com/abstract=2652446; Christopher S. Yoo, Cyber Espionage or Cyberwar?: International Law, Domestic Law, and Self-Protective Measures, In Cyberwar: Law and Ethics for Virtual Conflicts (Jens David Ohlin, Kevin Govern, Claire Finkelstein, eds., 2015), https://ssrn.com/abstract=2596634.

[24] *Id., citing* Peter C. Ormerod & Lawrence J. Trautman, *A Descriptive Analysis of the Fourth Amendment and the Third-Party Doctrine in the Digital Age*, 28 ALBANY L.J. SCI. & TECH. 73 (2018), https://ssrn.com/abstract=3005714.

[25] *See* Trey Herr, Bruce Schneier & Christopher Morris, Taking Stock: Estimating Vulnerability Rediscovery, Belfer Cyber Security Project White Paper Series (2017), https://ssrn.com/abstract=2928758.

[26] See NIST, *Securing Small Business and Home Internet of Things (IoT) Devices: Mitigating Network-Based Attacks By Using Manufactured Usage Description,* Nat'l Cybersecurity Center of Excellence (Feb. 2019).

EARLY DRAFT-COMMENTS WELCOME-1/27/2020 9:12 PM

use the results to better inform decision making, alter the physical environment, and anticipate future events. While the full scope of IoT is not precisely defined, it is clearly vast. Every sector has its own types of IoT devices, such as specialized hospital equipment in the healthcare sector and smart road technologies in the transportation sector, and there is a large number of enterprise IoT devices that every sector can use. Versions of nearly every consumer electronics device, many of which are also present in organizations' facilities, have become connected IoT devices—kitchen appliances, thermostats, home security cameras, door locks, light bulbs, and TVs.[27]

During 2019, Giaretta, Dragoni and Massacci provide the following summary of the IoT environment:

> According to Gartner Hype Cycle for Emerging Technologies, Internet of Things (IoT) surpassed the so-called peak of disillusion, headed to an established role within society. But all the problems are far from being solved and, the more pervasive the IoT becomes, the harder it is to manage. In particular, IoT security is one of the biggest cybersecurity challenges, and one of its most embarrassing failures. Traditional cybersecurity solutions have proven to be ineffective for IoT due to a number of technical and operational challenges. First, IoT devices are highly heterogeneous, with huge differences across tiers, languages, OSes, and networks. Also, the IoT lacks a common security framework, and standards are still not settled. Often times, security is not a manufacturers' (nor IT adminis') core competency, and may not be even considered part of the IoT product development process.[28]

A recent Google search identified consumer products such as a front door IoT camera monitor having a sales price point of US$34.95; and a baby monitor offering "Pet Camera Wireless IP Security WiFi Surveillance Camera with Cloud Storage Two Way Audio Pan/Tilt/Zoom Night Vision Motion Detect Remote Control for Home/Shop/

---

[27] Katie Boeckl, Michael Fagan, William Fisher, Naomi Lefkovitz, Katerina N. Megas, Ellen Nadeau, Danna Gabel O'Rourke, Ben Piccarreta & Karen Scarfone, Considerations for Managing Internet of Things (IoT) Cybersecurity and Privacy Risks, NISTR 8228 iv (June 2019), https://nvlpubs.nist.gov/nistpubs/ir/2019/NIST.IR.8228.pdf.
[28] *See* Giaretta, et al., *supra* note 3.

Office," priced at US$39.95.[29] Exhibit 1 provides an example of just one of the many IoT devices that connect to the Internet. Dozens of IoT sensors may be found in a typical smart home, "whereas industrial applications can scale up to hundreds of IoT devices. This introduces a number of problems, such as maintaining and monitoring the IoT devices, allowing and disallowing communication protocols, and overseeing what kind of information can be shared under defined conditions."[30]

Exhibit 1
Example of Internet Connectivity Device[31]

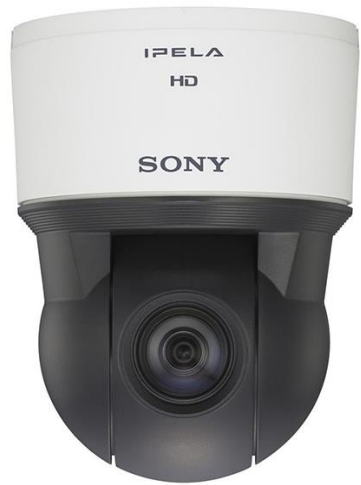

The Promise of The Internet of Things

Kris Alexander, CTO of Akamai Technologies notes the promise of the Internet of Things or the Internet of Everything:

> The promise of the IoT/IoE is that devices can now connect together (and with people) to enable new actions – to do something they couldn't before; like to warn you when your resting heart rate is too high, or learn how cool

---

[29] Recent Google search for 'IoT camera.'
[30] *See* Giaretta, et al., *supra* note 3.
[31] Recent Google search for 'IoT camera.'

you like your house and when you get home, and adjust the temperature before you get there.[32]

According to Bruce Sinclair, system integration engineers "look at IoT technology as a networking stack, which is, in a sense simply a protocol map."[33] (See Exhibit 2). Mr. Sinclair then hastens to add, "Mapping protocols from where the sensor data comes in, to the application is the absolute wrong way to look at the tech–at least for business. This is plumbing and not where the value originates."[34]

Exhibit 2
Network Engineer's View of IoT[35]

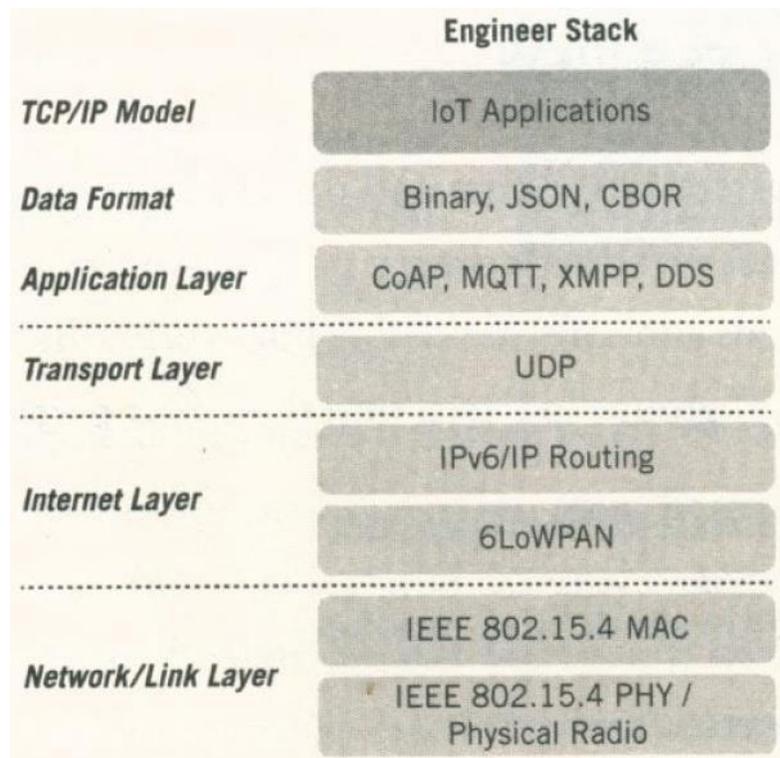

---

[32] Kris Alexander, CTO, Akamai Technologies; https://juniper-networks.cioreview.com/cxoinsight/the-promise-and-challenges-of-an-internet-of-things-iot-world-nid-4769-cid-73.html
[33] *See* Sinclair, *supra* note 11 at 5.
[34] *Id.*
[35] *See* Sinclair, *supra* note 11 at 5.

Writing from a business value perspective, Bruce Sinclair observes, "but plumbing is a means to an end; it is the way to get data from one place to another."[36] While Sinclair writes, "I don't look at IoT tech as a networking stack because it doesn't properly isolate and highlight where value is created;[37] mention of the networking stack lends value to our discussion here.

### III. THE CORPORATE GOVERNANCE CHALLENGE

> *"[T]he frequency and impact of cyber-attacks on our nation's private sector and government networks have increased dramatically in the past decade and are expected to continue to grow. We continue to see an increase in the scale and scope of reporting on malicious cyber activity that can be measured by the amount of corporate data stolen or deleted, personally identifiable information compromised, or remediation costs incurred by U.S. victims. Within the FBI, we are focused on the most dangerous malicious cyber activity: high-level intrusions by state-sponsored hackers and global organized crime syndicates, as well as other technically sophisticated attacks."*
>
> *Christopher Wray*
> *Director*
> *Federal Bureau of Investigation*
> *September 27, 2017*[38]

In the very briefest of terms, corporate officers and directors have two primary duties to shareholders: a *duty of loyalty* (no self dealing); and a duty of care (a duty to behave reasonably. The duty of care applies across director's and officers' myriad responsibilities, including handling the corporation's digital data. The duty of care is substantially procedural. During recent years, increased focus has been brought to bear on the responsibility of directors to ensure the data privacy of customers and users.[39] NIST observes:

> Many organizations are not necessarily aware they are using a large number of IoT devices. It is important that organizations understand their use of IoT because many IoT devices affect cybersecurity and privacy risks differently than conventional IT devices do. Once organizations are aware of their existing IoT usage and possible future usage, they need to understand how the characteristics of IoT affect managing cybersecurity and privacy risks, especially in terms of risk response—accepting, avoiding, mitigating, sharing, or transferring risk.[40]

---

[36] *Id.*
[37] *Id.*
[38] *See* Lawrence J. Trautman & Peter C. Ormerod, *WannaCry, Ransomware, and the Emerging Threat to Corporations*, 86(2) TENN. L. REV. 503, 507 (2019), http://ssrn.com/abstract=3238293, *citing* Current Threats to the Homeland: Hearings Before the S. Homeland Security and Govt. Affairs Comm, 115th Cong. (2017) (statement of Christopher Wray, Dir., Fed. Bureau of Investigation).

We now present a very brief discussion of the corporate duties of loyalty and care.

Duty of Loyalty

Under Delaware law, the duty of loyalty requires "that there shall be no conflict between duty and self-interest."[41] The core concept of the fiduciary "duty of loyalty" has been described as:

> [t]he requirement that a director favor the corporation's interests over her own whenever those interests conflict. As with the duty of care, there is a duty of candor aspect to the duty of loyalty. Thus, whenever a director confronts a situation that involves a conflict between her personal interests and those of the corporation, courts will carefully scrutinize not only whether she has unfairly favored her personal interest in that transaction, but also whether she has been completely candid with the corporation and its shareholders.[42]

Conflicts of interest "do not per se result in a breach of the duty of loyalty. Rather, it is the manner in which an interested director handles a conflict and the processes invoked to ensure fairness to the corporation and its stockholders that will determine the propriety of the director's conduct…"[43] Generally, except in cases where a director has an undisclosed financial interest in the outcome of a major corporate purchase or contract decision, the duty of loyalty does not seem to require additional focus.

---

[39] *See* Lawrence J. Trautman, *Governance of the Facebook Privacy Crisis*, 20 PITT. J. TECH. L. & POL'L (2020), http://ssrn.com/abstract=3363002.

[40] *See* Boeckl, et. al., *supra* note 27 at iv.

[41] *See* Lawrence J. Trautman & Kara Altenbaumer-Price, *The Board's Responsibility for Information Technology Governance,* 29 J. MARSHALL J. COMP. & INFO. L. 313, 324, (2011), *citing Guth v. Loft,* A.2d 503, 510 (Del. 1939).

[42] *Id. citing* Charles R.T. O'Kelley and Robert B. Thompson, *CORPORATIONS AND OTHER BUSINESS ASSOCIATIONS: CASES AND MATERIALS* (Aspen, 5th ed. 2006).

Duty of Care

Trautman and Altenbaumer-Price have previously written that, "The duty of care for directors 'arises in both the discrete decision-making context and in the oversight and monitoring areas.'[44] Prior the landmark 1985 case *Smith v. Van Gorkom*,[45] absent accompanying disloyal acts, it was generally accepted that "courts had rarely found individual directors liable for breaching their duty of care."[46] Experienced and sophisticated directors in that case were not entitled to the protection of the business judgment rule in some cases because:

> the duty of care specifies the manner in which directors must discharge their legal responsibilities… includ[ing] electing, evaluating, and compensating corporate officers; reviewing and approving corporate strategy, budgets, and capital expenditures; monitoring internal financial information systems and financial reporting obligations, and complying with legal requirements; making distributions to shareholders; approving transactions not in the ordinary course of business; appointing members to committees and discharging committee assignments, including the important audit, compensation and nominating committees; and initiating changes to the certificate of incorporation and bylaws.[47]

---

[43] *Id. citing* Byron Egan, Director Duties: Process and Proof, TexasBarCLE Webcast: Corporate Minutes/ Director Duties (Oct. 23, 2008) (www.jw.com/site/jsp/publicationinfo.jsp?id=1044).

[44] *See* Trautman & Altenbaumer-Price, *supra* note 41 at 322, *citing* Lyman P.Q. Johnson and Mark A. Sides, *Corporate Governance and the Sarbanes-Oxley Act: The Sarbanes-Oxley Act and Fiduciary Duties,* 30 WM. MITCHELL L. REV. 1149, 1197 (2004) (*citing Citron v. Fairchild Camera & Instrument Corp.,* 569 A.2d 53, 66 (Del. 1989)); *Brehm v. Eisner,* 746 A.2d 244, 264 (Del. 2000) ("Due care in the decision making context is process due care only.").

[45] *See* Trautman & Altenbaumer-Price, *supra* note 41 at 322, *citing Smith v. Van Gorkom*, 488 A.2d 858 (Del.Supr. 1985). The Delaware Supreme Court found that the experienced and sophisticated directors of Trans Union Corporation were not entitled to the protection of the business judgment rule and had breached their fiduciary duty to their shareholders "(1) by their failure to inform themselves of all information reasonably available to them and relevant to their decision to recommend the Pritzker merger; and (2) by their failure to disclose all material information such as a reasonable shareholder would consider important in deciding whether to approve the Pritzker offer." *Id.* at 888; see also *See* Peter V. Letsou, *Cases and Materials on Corporate Mergers and Acquisitions* n21 at 643 (2006) (observing "Trans Union's five 'inside' directors had backgrounds in law and accounting, 116 years of collective employment by the company and 68 years of combined experience on its Board. Trans Union's five 'outside' directors included four chief executives of major corporations and an economist who was a former dean of a major school of business and chancellor of a university. The 'outside' directors had 78 years of combined experience as chief executive officers of major corporations and 50 years of cumulative experience of Trans Union. Thus, defendants argue that the Board was eminently qualified to reach an informed judgment on the proposed 'sale' of Trans Union notwithstanding their lack of any advance notice on the proposal, the shortness of their deliberation, and their determination not to consult with their investment banker or to obtain a fairness opinion.").

Duty to Provide Data Security

The broad Duty of Care includes a duty to provide data security. Professors Trautman and Ormerod write:

> The duty of care applies across directors' and officers myriad responsibilities, including handling the corporation's digital data. There is, therefore, an emerging specific application of the duty of care as related to information technology: the duty to secure data. The applicable standard of care requires directors "to provide 'reasonable' or 'appropriate' physical, technical, and administrative security measures to ensure the confidentiality, integrity, and availability of corporate data."[48]

---

[46] *See* Trautman & Altenbaumer-Price, *supra* note 41 at 322, *citing* Jacqueline M. Veneziani, *Note & Comment: Causation and Injury in Corporate Control Transactions: Cede & Co. v. Technicolor, Inc.,* 69 WASH. L. REV. 1167, 1194 n.3 (1994) ("Before *Van Gorkom* was decided, one commentator had stated that '[t]he search for cases in which directors… have been held liable in derivative suits for negligence uncomplicated by self-dealing is a search for a very small number of needles in a very large haystack.' Joseph W. Bishop, Jr., *Sitting Ducks and Decoy Ducks: New Trends in the Indemnification of Corporate Directors and Officers,* 77 YALE L.J. 1078, 1099 (1968). *But see* Norwood P. Beveridge, Jr., *supra* note _ at 945-46 (disputing Prof. Bishop's statement and noting that there are actually many cases upholding duty of care violations)."

[47] *See* Lawrence J. Trautman & Peter C. Ormerod, *Corporate Directors' and Officers' Cybersecurity Standard of Care: The Yahoo Data Breach,* 66 AM. U. L. REV. 1231, 1232 (2017), [Hereinafter "Yahoo Data Breach"]http://ssrn.com/abstract=2883607, *citing* Lyman P.Q. Johnson and Mark A. Sides, *Corporate Governance and the Sarbanes-Oxley Act: The Sarbanes-Oxley Act and Fiduciary Duties,* 30 WM. MITCHELL L. REV. 1149, 1197 (2004) *citing Citron v. Fairchild Camera & Instrument Corp.,* 569 A.2d 53, 66 (Del. 1989); *Brehm v. Eisner,* 746 A.2d 244, 264 (Del. 2000) ("Due care in the decision-making context is process due care only.").

There is not, however, a single source—such as a comprehensive federal statute or regulation—that imposes a duty to provide data security. Rather, corporate legal obligations to implement data security systems are "set forth in an ever-expanding patchwork" of state, federal, and international statutes; regulations; enforcement actions; and common law duties, including "contractual commitments, and other expressed and implied obligations to provide 'reasonable' or 'appropriate' security for corporate data."[49]

Leadership at the Top

Any effective enterprise program to defend against cyberattack requires a commitment from top management to clearly communicate that good cyber hygiene is important and a top priority, not just empty rhetoric. Adequate resources must be provided in every organization if realistic progress against cyberthreat is to be made. Cyber attack and data theft is a real threat for all organizations: entrepreneurial start-ups;[50] non-profits;[51] municipalities;[52] educational institutions;[53] as well as large corporate entities.[54]

---

[48] THOMAS J. SMEDINGHOFF, INFORMATION SECURITY LAW: THE EMERGING STANDARD FOR CORPORATE COMPLIANCE 29 (2008).
[49] *Id.* at 29. *See also* Lawrence J. Trautman, *Cybersecurity: What About U.S. Policy?,* 2015 U. ILL. J. L. TECH. & POL'Y 341 (2015), http://ssrn.com/abstract=2548561; Lawrence J. Trautman, *Congressional Cybersecurity Oversight: Who's Who & How It Works,* 5 J. L. & CYBER WARFARE 147 (2016), http://ssrn.com/abstract=2638448.
[50] *See* Lawrence J. Trautman, Anthony "Tony" Luppino & Malika S. Simmons, *Some Key Things U.S. Entrepreneurs Need to Know About The Law and Lawyers,* 46 TEXAS J. BUS. L. 155 (2016), http://ssrn.com/abstract=2606808.
[51] *See* Lawrence J. Trautman & Janet Ford, *Nonprofit Governance: The Basics*, 52 AKRON L. REV. 971 (2018), https://ssrn.com/abstract=3133818.
[52] Lawrence J. Trautman, *Cybersecurity: What About U.S. Policy?,* 2015 U. ILL. J. L. TECH. & POL'Y 341 (2015), http://ssrn.com/abstract=2548561.
[53] David D. Schein & Lawrence J. Trautman, The Dark Web and Employer Liability 18(1) COLO. TECH. L.J. (2019), http://ssrn.com/abstract=3251479.
[54] *See* Lawrence J. Trautman, *The Board's Responsibility for Crisis Governance,* 13 HASTINGS BUS. L.J. 275 (2017), http://ssrn.com/abstract=2623219.

Board Talent and Experience

Corporate board nominating committees are challenged with the task of locating and recruiting directors who have the skills and experience necessary to govern data and information systems risk.[55] Very few directors have a background in computer science or electrical engineering. As a result, many boards are frustrated by the task of governing something they know very little about.[56]

Audit or Risk Committee Domain

Corporate boards operate through committees. Ensuring that the enterprise has a robust defense against cyberattack is handled by the board's audit committee in many organizations.[57] Other boards have created a risk committee with data security being an increasing focus for many organizations.

## IV. POTENTIAL IoT THREATS

Author Trautman has now spoken about IoT vulnerabilities on numerous occasions and has found his audiences particularly receptive to the following description of how data security threats are contained in IoT use. He asks his audience to "please close your eyes and imagine you are walking on a dark beach at night. As you walk you detect minor pricks to your feet (it feels like maybe a mosquito bite, but it isn't). After a while you realize that your legs have become numb."[58] Unknown to you, "the beach is covered with billions of hypodermic needles, all containing a localized anesthetic numbing agent like Novocaine (some contaminated with virus−think Ebola or HIV). This approximates the risk each of us is experiencing with IoT. You may open your eyes now."[59] Director of National Intelligence Daniel R. Coats in his prepared remarks before the Senate Select Committee on Intelligence for the 2017 hearings on the Worldwide Threat Assessment of the U.S. Intelligence Community observes that

> The widespread incorporation of "smart" devices into everyday objects is changing how people and machines interact with each other and the world around them, often improving efficiency, convenience, and quality of life. Their deployment has also introduced vulnerabilities into both the infrastructure that they support and on which they rely, as well as the processes they guide. Cyber actors have already used IoT devices for distributed denial-of-service (DDoS) attacks, and we assess they will continue. In the future, state and non-state actors will likely use IoT devices to support intelligence operations or domestic security or to access or attack targeted computer networks.[60]

---

[55] *See* Lawrence J. Trautman, *The Matrix: The Board's Responsibility for Director Selection and Recruitment,* 11 FLA. ST. U. BUS. REV. 75 (2012), http://www.ssrn.com/abstract=1998489.

[56] *See* Lawrence J. Trautman, *Who Qualifies as an Audit Committee Financial Expert Under SEC Regulations and NYSE Rules?,* 11 DEPAUL BUS. & COMM. L. J. 205 (2013), http://www.ssrn.com/abstract=2137747.

[57] *See* Lawrence J. Trautman, *Who Qualifies as an Audit Committee Financial Expert Under SEC Regulations and NYSE Rules?,* 11 DEPAUL BUS. & COMM. L. J. 205 (2013), http://www.ssrn.com/abstract=2137747.

Another way to think about IoT security vulnerabilities is to consider how you take the time to see that your home or apartment, containing your valuable possessions, is securely locked when you leave it unattended. Here, just as with your valuable personal data assets, the use of poorly secured IoT devices is roughly equivalent to locking the front door of your home and leaving the back door wide open. Consider the extent to which your home and family are vulnerable to unsecure IoT devices. Professors Streiff, Das and Cannon write, "The sensor capabilities of IoT toys along with other critical data, including location information, possess significant risk for malicious activity. Even unsophisticated attacks leading to location leakage can be problematic for vulnerable populations such as children. For parents, these are risks for which they are generally unprepared."[61] Courtesy of Bruce Sinclair, a graphic illustration of IoT threat vectors is presented as Exhibit 3.

---

[58] Lawrence J. Trautman, Remarks presented at the Seventh Annual Conference on Governance of Emerging Technologies, Sandra Day O'Connor College of Law, Arizona State University, May 22 & 23, 2019 (on file with authors).
[59] *Id.*
[60] Worldwide Threat Assessment of the U.S. Intelligence Community, Before the S. Select Comm. on Intelligence (115th Cong. 1 (2017) (statement of Daniel R. Coats, Director of National Intelligence).

Exhibit 3
Attack Vectors in IoT[62]

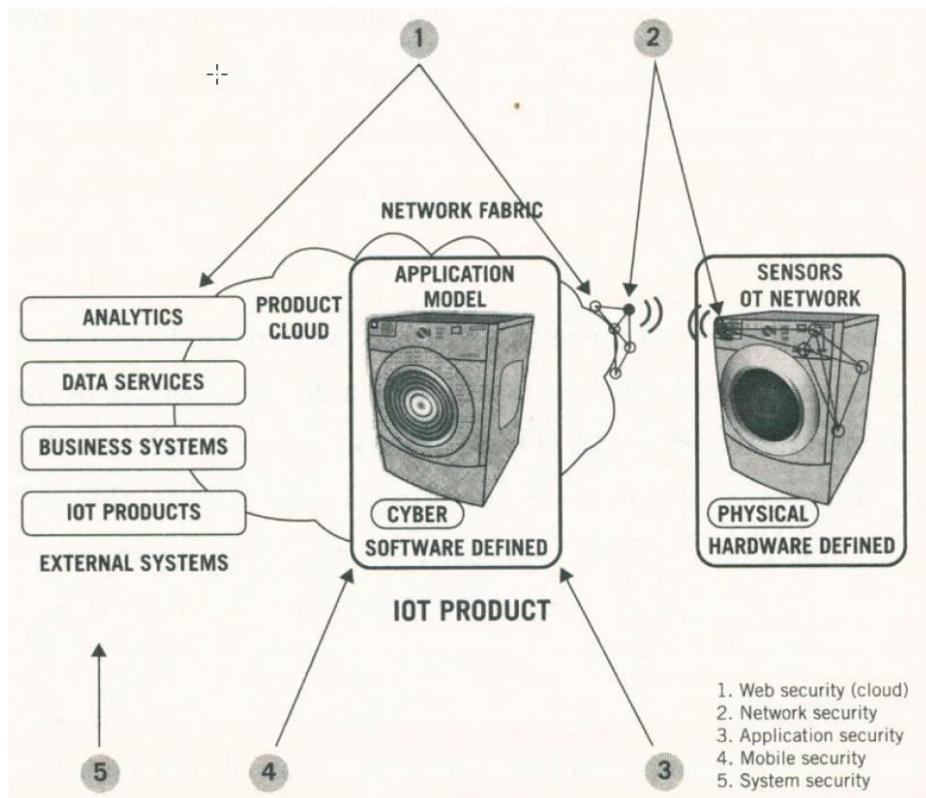

---

[61] *See* Joshua Streiff, Sanchari Das & Joshua Cannon, *Overpowered and Underprotected Toys Empowering Parents With Tools to Protect Their Children,* IEEE Humans and Cybersecurity Workshop (HACS 2019), https://ssrn.com/abstract=3509530.
[62] *See* Sinclair, *supra* note 11 at 244.

Data Breaches Continue

By now, every reader should be aware of the continued threat posed by inadequate data security.[63] A comprehensive discussion about the history, nature, and current threat profile of data breaches is beyond the scope of this single Article. However, we have chosen to mention and describe briefly the following data breaches: Target (December 2013); Yahoo (2013, but not reported until several years later); Equifax (2017); Office of Personnel Management (June 2015); Marriott Hotels (January 2019); and Capital One Financial Corp. (July 2019). These representative and widespread breaches have been chosen, in part, because one of your authors believes he was a victim of each. In addition, we are including a brief discussion about what is now known regarding the breach of at least 106 million card applicants of Capital One Financial Corp., first reported during July 2019. For historical reference, Exhibit 4 presents "Reported Incidents of Loss, Theft or Exposure of Personally Identifiable Information (PII)" for the period covering 2014 thru year-end 2018.

---

[63] *See* Julia Carpenter & Bouree Lam, *Consumers Feel Breach Fatigue,* WALL ST. J., Aug. 5, 2019 at B4.

Exhibit 4 [UPDATE]
Reported Incidents of Loss, Theft or Exposure of
Personally Identifiable Information (PII)[64]

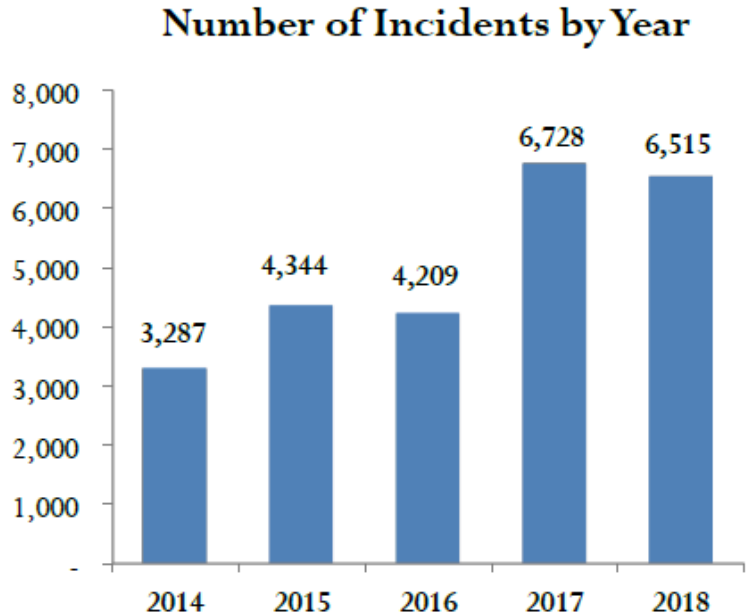

Exhibit 5 presents a list of the top 11 all-time Breaches as of July, 2019.

Exhibit 5
Top 11 All-time Breaches as of July, 2019[65]

1. **YAHOO.** Reported breach of 3 billion records on December 14, 2016.
2. **DU CALLER GROUP (China).** 2 billion records: customer names, addresses inappropriately made accessible in public directory, on May 13, 2017.
3. **RIVER CITY MEDIA (US).** 1.3 billion records: names; addresses; IP addresses; email addresses; undisclosed number of financial documents; chat logs and backup; exposed by faulty rsync backup, March 3, 2017.
4. **NETEASE, INC, dba 163.com (China).** Reports 1.2 billion records stolen by hackers (emails, addresses, and passwords) and offered for sale on the dark web, January 25, 2017.

---

[64] Footnote to follow when table is updated in February, 2020.
[65] *See* Cyber Risk Analytics, 2019 MidYear QuickView Data Breach Report, RiskBased Security 12 (Aug. 2019), https://pages.riskbasedsecurity.com/hubfs/Reports/2019/2019%20MidYear%20Data%20Breach%20QuickView%20Report.pdf.

5. **UNKNOWN (India).** Village-level enterprise operators sell access to the Aadhaar database, Jan.3, 2018. 1,190 million names, addresses, email addresses, dates of birth, phone numbers, fax numbers, genders, IP addresses, and photographs of Indian citizens… January 3, 2018.
6. **VERIFICATIONS.IO (Estonia).** 982 million names addresses, email addresses, dates of birth, phone numbers, fax numbers, genders, IP addresses, personal mortgage amounts, and FTP server credentials, exposed on the Internet due to a misconfigured database. March 7, 2019.
7. **FIRST AMERICAN FINANCIAL CORPORATION (U.S.)** Approximately 885 million real estate closing transaction records containing names, Social Security numbers, phone numbers, email and physical addresses, driver's license images, banking details, and mortgage lender names and loan numbers exposed on the Internet due to IDOR flow.
8. **UNKNOWN (Netherlands).** Breach of 711 million records: email addresses; passwords; credentials exposed on the Internet due to misconfigured database, January 3, 2017.
9. **CULTURA COLECTIVA (Mexico).** 540 Million Facebook user IDs, account names, comments, and likes exposed on the Internet due to a misconfigured database.
10. **YAHOO (US).** Breach involving 500 million records: user names; email addresses; phone numbers; dates of birth; hashed passwords and security questions and associated answers, September 22, 2016.[66]

Author Trautman believes it is likely that his household has been negatively impacted by each of the following data breaches: Target (2013); Yahoo (2013); Equifax (2017); Office of Personnel Management (June 2015); Marriott (January 2019); and Capital One Financial Corp. (July 2019).

Target (2013)

U.S. retailer Target reported that the "information stolen between November 27 and December 15, 2013 included personal information of as many as 70 million people--- more than the 40 million the company originally estimated."[67] Several years later, Target agreed to "pay $18.5 million to 47 states and the District of Columbia as part of a settlement with state attorneys general."[68] In total, Target reports that the estimated cost of the breach approximates $300 million.[69]

---

[66] *Id.*

Yahoo (2013)

As shown in Exhibit 2, Top 11 All-time Breaches as of December 31, 2018, Yahoo! Inc. has the distinction of: (1) being responsible for several of the largest U.S. data breaches; and (2) being among the slowest to inform consumers of these breaches.[70] As reported by professors Trautman and Ormerod, On September 22, 2016 Yahoo! Inc. announced that a data breach and theft of information from over 500 million user accounts had taken place during 2014. It now appears that this theft included names, birthdays, telephone numbers, email addresses, "hashed passwords (the vast majority with bcrypt) and, in some cases, encrypted or unencrypted security questions and answers."[71] This 2014 theft represents the largest data breach ever at the time it was announced.[72] Yahoo further disclosed their belief that the stolen data "did not include unprotected passwords, payment card data, or bank account information."[73] Just two months before Yahoo disclosed its 2014 data breach, a proposed sale of the company's core business to Verizon Communications was announced.[74] Yahoo then announced, during mid-December 2016, "that another 1 billion customer accounts had been compromised during 2013, establishing a new record for the largest data breach ever."[75]

---

[67] *See* Maggie McGrath, *Target Data Breach Spilled Info On As Many As 70 Million Customers,* FORBES (Jan. 10, 2014), https://www.forbes.com/sites/maggiemcgrath/2014/01/10/target-data-breach-spilled-info-on-as-many-as-70-million-customers/#6343ee5ae795.
[68] *See* Rachel Abrams, *Target to Pay $18.5 Million to 47 States in Security Breach Settlement,* N.Y. TIMES, May 23, 2017, https://www.nytimes.com/2017/05/23/business/target-security-breach-settlement.html.
[69] *See* Vincent Lynch, *Cost of 2013 Target Data Breach Nears $300 Million,* HashedOut, May 26, 2017.
[70] *See* Yahoo Data Breach. *supra* note 47 at 1233.
[71] *Id. citing* Press Release, Yahoo! Inc., An Important Message to Yahoo Users on Security (Sept. 22, 2016), https://investor.yahoo.net/releasedetail.cfm?releaseid=990570.
[72] *See* Nicole Perlroth, Yahoo Says Hackers Stole Data on 500 Million Users in 2014, N. Y. TIMES, Seot 22, 2016, https://www.nytimes.com/2016/09/23/technology/yahoo-hackers.html?_r=0.

Equifax (2017)

On September 7, 2017, global credit reporting agency Equifax "announced that its consumer information had been compromised as a result of a 'cybersecurity incident."[76] McKay Smith and Garrett Mulrain report that "this incident resulted in the loss of the personally identifiable information (PII) of 143 million American consumers, or nearly 45 percent of the American population."[77] How does something like this happen? It appears, "The Department of Homeland Security alerted Equifax officials on March 8, 2017 that they needed to fix a critical security vulnerability in their software. Company officials disseminated the alert internally but failed to manually patch the application. This single point of failure would prove to be catastrophic."[78]

---

[73] *Id.*
[74] *See* Yahoo Data Breach. *supra* note 47.
[75] *Id.* at 1234, *citing* Robert McMillan, Ryan Knutson & Deepa Seetharaman, Yahoo Discloses New Breach of 1 Billion User Accounts, WALL ST. J., Dec 15, 2016, https://www.wsj.com/articles/yahoo-discloses-new-breach-of-1-billion-user-accounts-1481753131.
[76] *See* McKay Smith & Garrett Mulrain, *Equi-Failure: The National Security Implications of the Equifax Hack and a Critical Proposal for Reform,* 9 J. NAT'L SEC. L. & POL'Y 549 (2018); *See also* Scott J. Shackelford & Austin E. Brady, Is It Time For A National Cybersecurity Safety Board? Examining The Policy Implications and Political Pushback (unpub ms) (2019).
[77] *Id. citing* Spencer Kimball & Liz Moyer, *Equifax Data Breach May Affect 2.5 Million More Consumers than Originally Stated,* CNBC (Oct. 2, 2017), https://www.cnbc.com/2017/10/02/equifax-2-point-5-million-more-consumers-may-be-affected-by-data-breach-than-originally-stated.html.
[78] *See* Smith & Mulrain, *supra* note 76 at 555.

### Office of Personnel Management (June 2015)

The June 2015 breach containing some of America's most sensitive information at the U.S. Office of Personnel constitutes a particular threat to American national security because the data stolen "was significant in that it specifically targeted security clearance information for the federal workforce… The catastrophic harm may… occur at some point in the future, to include 'the ability to blackmail, shame, or otherwise coerce public officials.'"[79]

### Marriott (January 2019)

As disclosed on November 30, 2018 and in Exhibit 3, hackers were able to compromise the hotel chain's loyalty program database, exposing 383 million records: names; addresses; reservation details; and passport numbers.[80] According to *The Washington Post*, "hackers have had access to the reservation systems of many of its hotel chains for the past four years, a breach that exposed private details of up to 500 million customers while underscoring the private nature of records showing where and when people travel−and with whom."[81]

### Capital One Financial Corp. (July 2019)

The fifth largest credit card issuer in the United States, Capital One Financial Corp., reported the breach of at least 106 million records of card customers and applicants during late July 2019.[82] As an indication of the cost to corporations of these data breaches, common stock share values for Capital One Financial Corp. closed down 5.9% of the day of announcement, July 30, 2019.[83] *The Wall Street Journal* reports, the arrest of "Paige A. Thompson, a former employee at Amazon.com Inc's cloud-computing unit."[84] *The Wall Street Journal* story continues to report that "the largest-ever bank-data heists appeared to have exploited a vulnerability in the cloud that security experts have warned about for years."[85]

---

[79] *Id.* at 563.
[80] *See* Taylor Telford & Craig Timberg, *Marriott Discloses Massive Data Breach Affecting Up to 500 Million Guests,* WASH. POST, Nov. 30, 2018, https://www.washingtonpost.com/business/2018/11/30/marriott-discloses-massive-data-breach-impacting-million-guests/?noredirect=on.
[81] *Id.*

Ransomware

Are the billions of IoT sensors the gateway vehicle for ransomware attacks? During 2018, professors Trautman and Ormerod provided an extensive account of the history and evolution of ransomware,[86] and we will not duplicate that effort here. The Federal Bureau of Investigation (FBI) defines ransomware as:

> a type of malware installed on a computer or server that encrypts the files, making them inaccessible until a specified ransom is paid. Ransomware is typically installed when a user clicks on a malicious link, opens a file in an e-mail that installs the malware, or through drive-by downloads (which does not require user-installation) from a compromised Web site.[87]

The FBI further states that, "hospitals, school districts, state and local governments, law enforcement agencies, small businesses, large businesses−these are just some of the entities impacted by ransomware, an insidious type of malware that encrypts, or locks, valuable digital files and demands a ransom to release them."[88] The U.S. Department of Homeland Security's National Cybersecurity and Communications Integration Center (NCCIC) warns, "Ransomware not only targets home users; businesses can also become infected with ransomware, leading to negative consequences, including: temporary or permanent loss of sensitive or proprietary information, disruption to regular operations, financial losses incurred to restore systems and files, and potential harm to an organization's reputation."[89] Consider that, "the inability to access the important data these kinds of organizations keep can be catastrophic in terms of the loss of sensitive or proprietary information, the disruption to regular operations, financial losses incurred to restore systems and files, and the potential harm to an organization's reputation."[90] The FBI warns, "in a ransomware attack, victims−upon seeing an e-mail addressed to them−will open it and may click on an attachment that appears legitimate, like an invoice or an electronic fax, but which actually contains the malicious ransomware code."[91] Alternatively, "the e-mail might contain a legitimate-looking URL, but when a victim clicks on it, they are directed to a website that infects their computer with malicious software."[92] In addition:

> Once the infection is present, the malware begins encrypting files and folders on local drives, any attached drives, backup drives, and potentially other computers on the same network that the victim computer is attached to. Users and organizations are generally not aware they have been infected until they can no longer access their data or until they begin to see computer messages advising them of the attack and demands for a ransom payment in exchange for a decryption key. These messages include instructions on how to pay the ransom, usually with bitcoins because of the anonymity this virtual currency provides.[93]

---

[82] *See* Stacy Cowley & Nicole Perlroth, *One Gap in Bank's Armor, and a Hacker Slips In,* N.Y. TIMES, July 31, 2019 at A1; Nicole Hong, Liz Hoffman & AnnaMaria Andriotis, *Capital One Breach Affects 106 Million Card Applicants,* WALL ST. J., July 30, 2019 at A1.
[83] *See* Gunjan Banerji, *Capital One Shares Sink After Breach Is Disclosed,* WALL ST. J., July 31, 2019 at B12.
[84] *See* Robert McMillan, *Accused Hacker Exploited Vulnerability in Cloud,* WALL ST. J., Aug. 5, 2019 at A1; *See also* Dana Mattioli, Robert McMillan & Sebastian Herrera, *Hacking Suspect Left Trail of Clues Online,* WALL ST. J., July 31, 2019 at A1.
[85] *Id.*
[86] *See* Trautman & Ormerod, *supra* note 13.
[87] *See* Ransomware Victims Urged to Report Infections to Federal Law Enforcement, FBI Public Service Announcement, Alert No. I-091516-PSA, https://www.ic3.gov/media/2016/160915.aspx.

---

[88] *See* Cyber Crime: Key Priorities, Ransomware, https://www.fbi.gov/investigate/cyber (last viewed Aug. 24, 2018).
[89] *See* NCCIC Alert (TA17-132A), Indicators Associated With WannaCry Ransomware (June 7, 2018), https://www.us-cert.gov/ncas/alerts/TA17-132A.
[90] *See* Cyber Crime: Key Priorities, Ransomware, https://www.fbi.gov/investigate/cyber (last viewed Aug. 24, 2018).
[91] *Id.*
[92] *Id.*

Malicious and costly ransomware attacks continue daily, including and resulting in substantial disruption to the citizens of Atlanta,[94] Baltimore,[95] and many others.[96] New ransomware exploits are found constantly. For example, on January 22, 2020, security expert Ravi Gidwani reports, "a nasty and one-of-its-kind ransomware… one that uses Note.js framework, which enables it to infect Windows based OS."[97] This is significant because:

> Node.js is an open-source, cross-platform, JavaScript run-time environment that executes JavaScript code outside of a browser. It is built on the V8 JavaScript engine… Google's open source high- performance JavaScript and WebAssembly engine, written in C++. It is used in Chrome and in Node.js, among others. It implements ECMAScript and WebAssembly, and runs on Windows 7 or later, macOS 10.12+, and Linux systems that use x64, IA-32, ARM, or MIPS processors. V8 can run standalone, or can be embedded into any C++ application. Interestingly, users can easily get infected by this Nodera ransomware while browsing online, either by clicking on a malicious HTA file or when served as a malvertisement.[98]

Software engineer Tim Trautman believes this may be significant because, as a JavaScript framework, "Node has a very large community around it. Ransomware running on Node could increase the prevalence of Ransomware attacks by lowering the technical barrier of entry to a much larger / less technically-savy pool of software engineers."[99]

---

[93] *Id.*
[94] *See* Press Release, Atlanta U.S. Attorney Charges Iranian Nationals for City of Atlanta Ransomware Attack (Dec. 5, 2018), https://www.justice.gov/usao-ndga/pr/atlanta-us-attorney-charges-iranian-nationals-city-atlanta-ransomware-attack.
[95] *See* Niraj Chokshi, *Hackers Are Holding Baltimore Hostage: How They Struck and What's Next,* N.Y. TIMES, May 22, 2019, https://www.nytimes.com/2019/05/22/us/baltimore-ransomware.html.
[96] *See* Trautman & Ormerod, *supra* note 38.
[97] *See* Ravi Gidwani, *First Note.js-based Ransomware: Nodera,* Quickheal.com (Jan. 22, 2020), https://web.archive.org/web/20200122181519/https://blogs.quickheal.com/first-node-js-based-ransomware-nodera/.
[98] *Id.*

## V. MIRAI BOTNET

Akamai reports starting to track a strain of malware during June 2016 that targets Internet of Things (IoT) devices and home Internet routers.[100] Soon thereafter, this malware, under the name of Mirai, spread worldwide.[101] Journalist Elie Bursztein considers the Mirai attack particularly remarkable because "they were carried out via small, innocuous Internet-of-Things (IoT) devices like home routers, air-quality monitors, and personal surveillance cameras. At its peak, Mirai infected over 600,000 vulnerable IoT devices."[102] Journalist Bursztein explains, "At its core, Mirai is a self-propagating worm… a malicious program that replicates itself by finding, attacking and infecting vulnerable IoT devices."[103]

> The replication module is responsible for growing the botnet size by enslaving as many vulnerable IoT devices as possible. It accomplishes this by (randomly) scanning the entire Internet for viable targets and attacking. Once it compromises a vulnerable device, the module reports it to the C&C servers so it can be infected with the latest Mirai payload, as the diagram above illustrates.
> 
> To compromise devices, the initial version of Mirai relied exclusively on a fixed set of 64 well-known default login/password

---

[99] Email from Tim Trautman to Lawrence J. Trautman, 15:03 CST, Jan. 22, 2020 (on file with authors).
[100] *See* Akamai's [State of the Internet] / Security, Q3 2016 Report, 6 Akami Technologies, Inc. (2016), *https://www.akamai.com/us/en/multimedia/documents/state-of-the-internet/q3-2016-state-of-the-internet-security-report.pdf*.
[101] *Id.*
[102] *See* Elie Bursztein, *Inside the Infamous Mirai IoT Botnet: A Retrospective Analysis,* Cloudflare Blog (Dec. 14, 2017), *https://blog.cloudflare.com/inside-mirai-the-infamous-iot-botnet-a-retrospective-analysis/*.
[103] *Id.*

combinations commonly used by IoT devices. While this attack was very low tech, it proved extremely effective…[104]

Akamai observes that use of IoT devices and other capabilities usually not found in botnets make Mirai "truly exceptional… specifically Generic Routing Encapsulation (GRE) based attacks, varying levels of attack traffic customization, and telnet scanning. In addition, it generates its attacks directly…. Due to the public release of the source code… we're likely to see new, more-capable variants of Mirai in the near future."[105] In addition,

> Mirai is a botnet that would not exist if more networks practiced basic hygiene, such as blocking insecure protocols by default. This is not new—we've seen similar network hygiene issues as the source of infection in the Brobot attacks of 2011 and 2012. The botnet spreads like a worm, using telnet and more than 60 default username and password combinations to scan the Internet for additional systems to infect. The majority of these systems appear to be Digital Video Recorders (DVRs), ip-enabled surveillance cameras, and consumer routers. Once a system is infected, it connects to the command and control (C2) structure of the botnet, then continues scanning for other vulnerable systems while waiting for attack commands.[106]

During May 2018, the U.S. Departments of Commerce and Homeland Security jointly published, *A Report to the President on Enhancing the Resilience of the Internet and Communications Ecosystem Against Botnets and Other Automated, Distributed Threats.*[107] This publication was the result of consultations with interested agencies, including: "the Departments of Defense, Justice, and State, the Federal Bureau of Investigation, the sector-specific agencies, the Federal Communications Commission and

---

[104] *Id.*
[105] *See* Akamai, *supra* note 100 at 15.
[106] *Id.*
[107] *A Report to the President on Enhancing the Resilience of the Internet and Communications Ecosystem Against Botnets and Other Automated, Distributed Threats,* U.S. Departments of Commerce and Homeland Security (May 22, 2018).

Federal Trade Commission."[108] As a result of the rapid growth in the IoT devices, the Report notes that distributed denial of service "DDoS attacks have grown in size to more than one terabit per second, far outstripping expected size and excess capacity. As a result, recovery time from these types of attacks may be too slow, particularly when mission-critical services are involved."[109] These automated and distributed attacks (e.g., botnets), "are used for a variety of malicious activities… that overwhelm networked resources, sending massive quantities of spam, disseminating keylogger and other malware, ransomware attacks distributed by botnets that hold systems and data hostage."[110] The Report further states, "Traditional DDoS mitigation techniques, such as network providers building in excess capacity to absorb the effects of botnets, are designed to protect against botnets of an anticipated size… [but] were not designed to remedy other classes of malicious activities facilitated by botnets, such as ransomware or computational propaganda."[111] And:

> The DDoS attacks launched from the Mirai botnet in the fall of 2016, for example, reached a level of sustained traffic that overwhelmed many common DDoS mitigation tools and services, and even disrupted a Domain Name System (DNS) service that was a commonly used component in many DDoS mitigation strategies. This attack also highlighted the growing insecurities in—and threats from— consumer-grade IoT devices. As a new technology, IoT devices are often built and deployed without important security features and practices in place. While the original Mirai variant was relatively simple, exploiting weak device passwords, more sophisticated botnets have followed; for example, the Reaper botnet uses known code vulnerabilities to exploit a long list of devices, and one of the largest DDoS attacks seen to date recently exploited a newly discovered vulnerability in the relatively obscure MemCacheD software. These examples clearly demonstrate the risks posed by botnets of this size and scope, as well as the expected innovation and increased scale and complexity of future attacks.[112]

---

[108] *Id.* at 3.
[109] *Id.* at 5.
[110] *Id.*
[111] *Id.*

In December 2016, investigative cyber reporter, Brian Krebs released a story noting that:

> New research published this week could provide plenty of fresh fodder for Mirai, a malware strain that enslaves poorly-secured Internet of Things (IoT) devices for use in powerful online attacks. Researchers in Austria have unearthed a pair of backdoor accounts in more than 80 different IP camera models made by Sony Corp. Separately, Israeli security experts have discovered trivially exploitable weaknesses in nearly a half-million white-labeled IP camera models that are not currently sought out by Mirai.[113]

## VI.    IoT THREAT VECTORS DURING TIMES OF CRISIS

Governance challenges during times of crisis has been the focus of considerable scholarship during recent years.[114] Before addressing IoT vulnerabilities in times of crisis, it is important to provide a working definition of IoT system vulnerability, and a brief discussion on the crisis context.

<u>IoT System Vulnerability</u>

As discussed earlier (in section II), IoT is a vast network of devices that are connected to the internet. These devices interact with each other over wireless connections to create, exchange and transfer data without human interaction. Abomhara defines IoT vulnerabilities as weaknesses in an IoT system or its design that allow an intruder to execute commands, access unauthorized data, and/or conduct denial-of-service attacks.[115] Vulnerabilities can be flaws in the IoT hardware or software, weaknesses in policies and procedures used in the IoT systems or misuse of the IoT systems by the users themselves.[116]

---

[112] *Id.* at 6.
[113] https://krebsonsecurity.com/2016/12/researchers-find-fresh-fodder-for-iot-attack-cannons/
[114] *See* Trautman, *Crisis Governance, supra* note 54.
[115] *See* Mohamed Abomhara & Geir M. Køien, *Cyber Security and the Internet of Things: Vulnerabilities, Threats, Intruders and Attacks,* 4(1) J. CYBER SEC. & MOBILITY, 65 (2015).

Crisis Context

The World Health Organization (WHO) defines a disaster as any unforeseen event that causes damage, destruction, ecological disruption, loss of human life, human suffering, deterioration of health and health services on a scale that requires response efforts extending beyond the affected community.[117] Every year natural disasters cause significant economic losses and social impacts. With the ever-increasing population and infrastructures, the world's exposure to disaster related hazards is growing. Disaster contexts are volatile and often result in non-routine actions.[118] In time of disasters, knowledge availability varies extremely as compared to normal situations and people sometimes have to improvise to accommodate the condition critical for responding to the crisis. At times, disaster managers have to make decisions based on little or no information. The main characteristics of disasters are unpredictability, availability of limited resources in impacted areas, and dynamic changes in the environment.[119] During disasters, the impacts on people and infrastructures cannot be accurately predicted.

Disaster management involves creating plans through which people can alleviate vulnerability to hazards and cope with disasters. In managing a disaster, stakeholders collocated or geographically distributed are required to collaborate in order to provide effective and efficient disaster relief. Information and communication technologies play crucial roles in every step of the disaster management lifecycle (preparedness, response, recovery, mitigation).[120] With recent technological advances, IoT has the potential to become one of the most important enabling technologies for disaster management and relief.[121] Sinha et al. identifies the following three major application areas of IoT in disaster management (i) disaster risk minimization and prevention - monitoring disaster events with satellite communication, designing early warning systems, using social media for situational awareness; (ii) disaster response - real-time communication for effective and timely relief operations; and (iii) disaster recovery - online search for missing person and fund management systems. IoT can also be used to plan preventive maintenance and repairs; evaluate whether structures can withstand a coming weather event while continuing normal operations and close unsafe assets.[122] The dynamic nature of a disaster context stresses upon the ability to make efficient and precise decisions in minimal time.

---

[116] *Id.*

[117] *See* Community emergency preparedness: a manual for managers and policy-makers, World Health Organization (WHO) (1999), http://whqlibdoc.who.int/publications/9241545194.pdf (accessed June 25, 2017).

[118] *See* Jeannette Sutton, Leysia Palen & Irina Shklovski, *Backchannels on the Front Lines: Emergent Uses of Social Media in the 2007 Southern California Wildfires,* Proceedings of the 5th International ISCRAM Conference, Washington, DC, USA. (2008).

[119] *See* ManzhuYu, Chaowei Yang, & Yun Li, *Big Data in Natural Disaster Management: A Review,* 8(5), GEOSCIENCES, 165 (2018).

---

[120] See Louis Ngamassi, Thiagarajan Ramakrishnan & Shasedur Rahman, *Use of Social Media for Disaster Management: A Prescriptive Framework,* 28(3) J. ORGANIZATIONAL & END USER COMPUTING (JOEUC) (2016); Louis Ngamassi, Thiagarajan Ramakrishnan & Shasedur Rahman, Examining the Role of Social Media in Disaster Management from an Attribution Theory Perspective. Proceedings of the 13th International Conference on Information Systems for Crisis Response and Management (ISCRAM), Rio de Janeiro, Brazil May 22-25, (2016); Louis Ngamassi, A. Malik & D. Ebert, Social Media Visual Analytic Toolkits for Disaster Management: A Review of the Literature. International Conference on Information Systems for Crisis Response and Management (ISCRAM), Albi, France May 21-24, (2017).

[121] *See* A. Sinha, Kumar, P., Rana, N. P., Islam, R., & Dwivedi, Y. K. *Impact of internet of things (IoT) in disaster management: a task-technology fit perspective,* 283(1-2) ANNALS OF OPERATIONS RESEARCH, 759-794 (2019).

[122] *See* A. Sinha, Kumar, P., Rana, N. P., Islam, R., & Dwivedi, Y. K. *Impact of internet of things (IoT) in disaster management: a task-technology fit perspective,* 283(1-2) ANNALS OF OPERATIONS RESEARCH, 759-794 (2019).

The IoT technology, having the potential for communicating instantaneous information updates, can be a key player for realizing dynamic workflow adaptations.[123]

IoT Vulnerability in Time of Crisis

While IoT has enormous potential for disaster management, it also comes with a number of challenges during time of crisis. Large scale natural disasters such as the South Asian Tsunami in 2004, the Hurricane Katrina in 2005, the Haiti earthquake in 2010, and the Hurricane Harvey in 2017 led to massive destruction of information and communication technology infrastructures and highlighted the vulnerabilities of IoT systems.[124] In time of crisis, some of the major IoT system vulnerabilities are related to system interoperability and system interconnectivity two important concepts in IoT paradigm.

IoT systems rely on the interoperability of the different objects that are interconnected. IoT objects interact smartly through the Internet with other devices. In time of crisis, some of these devices may likely got destroyed which may significantly hamper the effective functioning of the IoT system. Therefore, the more objects get connected through IoT system, the greater becomes the possibility of mayhem in times of crises. Moreover, IoT services such as process automation, device management, decision making, are usually hosted on cloud to allow users to access IoT devices anytime, anywhere. If the Internet infrastructure is destroyed during a disaster, IoT services will not be available to the users. Technological advances such as "Content-Centric

---

[123] *Id.*
[124] *See* Trautman, *Crisis Governance, supra* note 54.

Networking (CCN) has become a promising network paradigm that satisfies the requirements of fast packets delivery for emergency applications of IoT."[125]

## VII.     MANUFACTURED USAGE DESCRIPTION (MUD) METHODOLOGY

NIST observes, "Unfortunately, IoT devices often lack efficient and effective features for customers to use to help mitigate cybersecurity risks."[126] The NIST internal report 8259 warns:

> Consequently, some IoT devices are less easily secured using customers' existing methods because the cybersecurity features they expect may not be available on IoT devices or may function differently than is expected based on conventional IT devices. This means IoT device customers may have to select, implement, and manage additional or new cybersecurity controls or alter the controls they already have. However, new or tailored controls to sufficiently mitigate risks to the same level as before may not be available to all customers or implementable with all IoT devices. Compounding this problem, customers may not know they need to alter their existing IT processes to accommodate IoT. The result is many IoT devices are not secured properly, so attackers can more easily compromise them and use them to harm device customers and conduct additional nefarious acts (e.g., distributed denial of service [DDoS] attacks) against other organizations.[127]

As the lack of standardization across the IoT market has led to a series of cyber-attacks and interoperability issues, the Manufactured Usage Description (MUD) has been developed as the industry's initial solution.[128] The international standards body, Internet Engineering Task Force (IETF), moved the draft MUD standard into a quasi-accepted proposed standard RFC 8520 in March 2019.[129] In response to the release to RFC 8520, the National Institute of Standards and Technology (NIST) released a draft of NIST Special Publication (NIST SP) 1800-15: Securing Small-Business and Home Internet of Things (IoT) Devices – Mitigating Network-Based Attacks Using Manufacturer Usage Description (MUD).[130] NIST is the standards body – particularly for cybersecurity standards – for the US Government.[131] The document describes MUD and its purpose:

> The goal of the Internet Engineering Task Force's manufacturer usage description (MUD) architecture is for Internet of Things (IoT) devices to behave as intended by the manufacturers of the devices. This is done by providing a standard way for manufacturers to identify each device's type and to indicate the network communications that it requires to perform its intended function. When MUD is used, the network will automatically permit the IoT device to perform as intended, and the network will prohibit all other device behaviors.[132]

---

[125] *See* Fawaz Alassery, *Fast Packet Delivery Techniques for Urgent Packets in Emergency Applications of Internet of Things,* 11(3) INT'L J. COMP. NETWORKS & COMMUNICATIONS (May 2019), https://ssrn.com/abstract=3405251.
[126] *See* Fagan, et al., *supra* note 4 at vii.
[127] *Id.*
[128] Ryan McCauley, *The Internet of Things Needs Standardization−Here's Why,* govtech.com (Mar. 1, 2017), https://www.govtech.com/fs/The-Internet-of-Things-Needs-Standardization-Heres-Why.html.

MUD-capable IoT devices for use in homes and small businesses make it more difficult for malicious actors to exploit these IoT devices to mount DDoS attacks across the Internet.[133] NIST explains that "MUD provides a standard method for access control information to be available to network control devices."[134] Distributed Denial of Service

---

[129] Manufacturer Usage Description Specification, Internet Engineering Task Force (IETF), RFC8520 (Mar. 2019), https://tools.ietf.org/html/rfc8520.
[130] Donna Dodson, Tim Polk, Murugiah Souppaya, William C. Barker, Parisa Grayeli & Susan Symington, Securing Small-Business and Home Internet of Things (IoT) Devices: Mitigating Network-Based Attacks Using Manufacturer Usage Description (MUD), NIST Special Pub. 1800-15A (Prelim. Draft Nov. 2019), https://csrc.nist.gov/publications/detail/sp/1800-15/draft.
[131] Public Law 113–283. 'Federal Information Security Modernization Act of 2014'
[132] Donna Dodson, Tim Polk, Murugiah Souppaya, Yemi Fashina, Parisa Grayeli, Joshua Klosterman, Blaine Mulugeta, Mary Raguso, Susan Symington, Jaideep Singh, William C. Barker, Dean Coclin, Clint Wilson, Darshak Thakore, Mark Walker, Eliot Lear, Brian Weis, Tim Jones & Drew Cohen, Securing Small-Business and Home Internet of Things (IoT) Devices: Mitigating Network-Based Attacks Using Manufacturer Usage Description (MUD), NIST Special Pub. 1800-15 (Prelim. Draft Apr. 2019), https://www.nccoe.nist.gov/sites/default/files/library/sp1800/iot-ddos-nist-sp1800-15-preliminary-draft.pdf.
[133] *See* Manufacturer Usage Description Specification, *supra* note 129.
[134] Donna Dodson, Tim Polk, Murugiah Souppaya, William C. Barker, Parisa Grayeli & Susan Symington, Securing Small-Business and Home Internet of Things (IoT) Devices: Mitigating

(DDoS) attacks are thwarted "by prohibiting unauthorized traffic to and from IoT devices. Even if an IoT device becomes compromised, MUD prevents it from being used in any attack that would require the device to send traffic to an unauthorized destination."[135] Exhibit 6 provides a schematic of this security methodology.

Exhibit 6 [136]
Manufactured Usage Development (MUD) Methodology

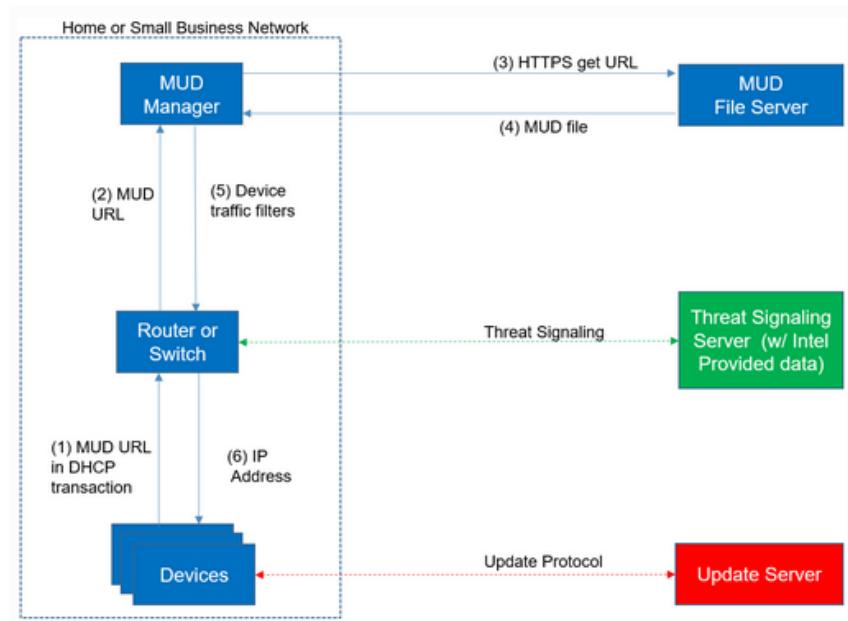

Description

A more detailed depiction of MUD is found in IETF's Manufacturer Usage Description Specification provides a description of what MUD fundamentally consists of: "three architectural building blocks: A URL that can be used to locate a description; The description itself, including how it is interpreted; and A means for local network management systems to retrieve the description.[137]

---

Network-Based Attacks Using Manufacturer Usage Description (MUD), NIST Special Pub. 1800-15A 1 (Prelim. Draft Nov. 2019), https://csrc.nist.gov/publications/detail/sp/1800-15/draft.
[135] *Id.*
[136] *See* NIST Draft 1800-15, *supra* note 130.

<u>Design</u>

The MUD intends to achieve several goals including:

- Substantially reduce the threat surface on a device to those communications intended by the manufacturer;
- Provide a means to scale network policies to the ever-increasing number of types of devices in the network;
- Provide a means to address at least some vulnerabilities in a way that is faster than the time it might take to update systems. This will be particularly true for systems that are no longer supported;
- Keep the cost of implementation of such a system to the bare minimum; and
- Provide a means of extensibility for manufacturers to express other device capabilities or requirements.[138]

These goals make the use of this framework practical while accomplishing a standardized level of security and use. However, the MUD design is not intended to:

- address network authorization of general purpose computers, as their manufacturers cannot envision a specific communication pattern to describe;
- In addition, even those devices that have a single or small number of uses might have very broad communication patterns. MUD on its own is not for them either;
- Although MUD can provide network administrators with some additional protection when device vulnerabilities exist, it will never replace the need for manufacturers to patch vulnerabilities;
- Finally, no matter what the manufacturer specifies in a MUD file, these are not directives, but suggestions. How they are instantiated locally will depend on many factors and will be ultimately up to the local network administrator, who must decide what is appropriate in a given circumstances.[139]

---

[137] *See* Manufacturer Usage Description Specification, *supra* note 129.
[138] *Id.*
[139] *Id.*

The Promise of Blockchain

In just about a decade blockchain technology has grown to be viewed with substantial promise for its potential to provide enhanced software security.[140] Just a few of the other promising applications include: smart contracts;[141] virtual currencies;[142] and numerous financial services functions, including execution and clearing.[143] Writing in 2018, Jianli Pan et al., observe:

> The emerging Internet of Things (IoT) is facing significant scalability and security challenges. On the one hand, IoT devices are 'weak' and need external assistance. Edge computing provides a promising direction addressing the deficiency of centralized cloud computing in scaling massive number of devices. On the other hand, IoT devices are also relatively 'vulnerable' facing malicious hackers due to resource constraints. The emerging blockchain and smart contracts technologies bring a series of new security features for IoT and edge computing.[144]

In an attempt to provide solutions to these issues, an edge-IoT framework named 'EdgeChain' is designed and prototyped by Jianli Pan et al., "based on blockchain and smart contracts. The core idea is to integrate a permissioned blockchain and the internal currency or 'coin' system to link the edge cloud resource pool with each IoT device' account and resource usage, and hence behavior of the IoT devices."[145] Consider:

> EdgeChain uses a credit-based resource management system to control how much resource IoT devices can obtain from edge servers, based on pre-defined rules on priority, application types and past behaviors. Smart contracts are used to enforce the rules and policies to regulate the IoT device behavior in a non-deniable and automated manner. All the IoT activities and transactions are recorded into blockchain for secure data logging and auditing. [Jianli Pan et al.] implement an EdgeChain prototype and conduct extensive experiments to evaluate the ideas. The results show that while gaining the security benefits of blockchain and smart contracts, the cost of integrating them into EdgeChain is within a reasonable and acceptable range.[146]

---

[140] *See* Lawrence J. Trautman & Mason J. Molesky, *A Primer for Blockchain,* 88(2) UMKC L. REV. 239 (2019), https://ssrn.com/abstract=3324660.
[141] *Id.*
[142] *See* Lawrence J. Trautman & Alvin Harrell, *Bitcoin Versus Regulated Payment Systems: What Gives?* 38 CARDOZO L. REV. 1041 (2017), http://ssrn.com/abstract=2730983; Lawrence J. Trautman, *Virtual Currencies: Bitcoin & What Now After Liberty Reserve, Silk Road, and Mt. Gox?,* 20 RICHMOND J. L. & TECH. 13 (2014), http://ssrn.com/abstract=2393537.
[143] *See* Lawrence J. Trautman, *Is Disruptive Blockchain Technology the Future of Financial Services?,* 69 CONSUMER FIN. L.Q. REP. 232 (2016), http://ssrn.com/abstract=2786186.
[144] *See* Jianli Pan, Jianyu Wang, Austin Hester, Ismail Alqerm, Yuanni Liu & Ying Zhao, *EdgeChain: An Edge-IoT Framework and Prototype Based on Blockchain and Smart Contracts,* arXiv:1806.06185v1 (June 16, 2018).
[145] *Id.*

The Jianli Pan et al. design schematic is constructed "Specifically, … [to] partially reference the Manufacturers Usage Description (MUD) files which list the activities and communications allowed for IoT devices. Such specifications contain input/output data type, requests of edge resources, MAC address, IP address, network port, communications protocol, and indication flags… each device registers a unique account address…"[147] Here's how it works: "Upon registration, the edge server will verify the [specified] information and take control of the modification rights of registration data. More parameters will be appended such as priority, coin balance, credit, and requests timestamp to benefit device management."[148] Pan et al. include several other attributes they defined in their registration database, "all the devices key information, value units, and examples, including: … account address; network port; input/output data; bandwidth request; CPU request; memory request; storage request; MAC address; priority*; coin balance*; credit*; isBlocked*; isRegistered*; [and] last request id*."[149] By way of explanation, Pan et al. observe, "Edge servers and IoT devices have different authorities to modify the registry. The attributes marked with an '*' can only be updated by the edge server. The other basic attributes are filled up during the first registration process initialized by IoT devices."[150]

---

[146] *Id.*
[147] *Id.*
[148] *Id.*
[149] Id.

### Importance of Consumer Education

Any engineering professor will tell you that human behavior and attitudes will play a determinative role in the success of any product design. NIST advises:

> Addressing the challenges of IoT cybersecurity necessitates educating IoT device customers on the differences in cybersecurity risks and risk mitigation for IoT versus conventional IT, as NIST has documented in Internal Report (IR) 8228, Considerations for Managing Internet of Things (IoT) Cybersecurity and Privacy Risks. The challenges also necessitate educating IoT device manufacturers on how to identify the cybersecurity features customers need IoT devices to have. This includes improving communications between manufacturers and customers regarding device cybersecurity features and related expectations.[151]

## VIII. RECENT DEVELOPMENTS

### NIST

The NIST continues to provide valuable research efforts, publications, and interface between governmental resources and industry. Professor Václav Janecek writes about the treasure trove of collected and created personal data from IoT devices, "whose management poses serious ethical and legal questions. Ownership of personal data underpins the issues revolving around data management and control, such as privacy, trust, and security, and it has also important implications for the future of the 'digital' economy and trade in data."[152]

---

[150] *Id.*
[151] *See* Fagan, et al., *supra* note 4 at vii.
[152] *See* Václav Janecek, *Ownership of Personal Data in the Internet of Things,* 34(5) COMP. L. & SEC. REV. 1039 (2018).

During January 2020, NIST released version 1.0 of its Privacy Framework.[153] This new tool for managing privacy risk contains "an overarching structure modeled on that of the widely used NIST Cybersecurity Framework and the two frameworks are designed to be complementary and also updated over time."[154] NIST observes that privacy interests, "includes information about specific individuals, such as their addresses or Social Security numbers, that a company might gather and use in the normal course of business… [requiring] an organization… take action to ensure [these data] are not misused in a way that could embarrass, endanger or compromise the customers."[155] While not a regulation or law, the Privacy Framework is "a voluntary tool that can help organizations manage privacy risk arising from their products and services, as well as demonstrate compliance with laws that may affect them."[156] Other recent NIST publications addressing IoT issues are available.[157]

As we approach the end of our discussion about cybersecurity and privacy risks, it is important to consider that data security "for an IoT device can [not] all be addressed within the devise itself. Every IoT device operates within a broader IoT environment where it interacts with other IoT and non-IoT devices, cloud-based services, people, and other components."[158]

---

[153] *See* NIST Privacy Framework: A Tool For Improving Privacy Through Enterprise Risk Management, Version 1.0, NIST (Jan. 16, 2020), https://www.nist.gov/system/files/documents/2020/01/16/NIST%20Privacy%20Framework_V1.0.pdf.
[154] *See* Press Release, NIST Releases Version 1.0 of Privacy Framework, NIST, U.S. Dept. of Commerce (Jan. 16, 2020), https://www.nist.gov/news-events/news/2020/01/nist-releases-version-10-privacy-framework.
[155] *Id.*
[156] *Id.*
[157] *See* Fagan, et al., *supra* note 4;
[158] *See* Boeckl et al., *supra* note 27 at 1.

<u>California Law SB. 327</u>

A new Californian law, starting in 2020, will require manufactures to implement "reasonable security feature[s]" on their connected devices or IoT.[159] This is one of the first regulations in the US to place information security requirements on all consumer products/devices and to place this burden on the manufacturers.

## IX. RECOMMENDATIONS

After considering a multitude of possible steps that can be taken, we have concluded that the goals and actions contained in the mid-2018, *A Report to the President on Enhancing the Resilience of the Internet and Communications Ecosystem Against Botnets and Other Automated, Distributed Threats*[160] is a good place to start. In relevant part, the Report states:

> These goals and actions aim to present a portfolio of mutually supportive actions that, if implemented, would dramatically improve the resilience of the ecosystem. The recommended actions include ongoing activities that should be continued or expanded, as well as new initiatives. No single investment or activity can mitigate all threats, but organized discussions and stakeholder feedback will allow us to further evaluate and prioritize these activities based on their expected return on investment and ability to measurably impact ecosystem resilience. We look to stakeholders across the ecosystem to work with government to implement the proposed activities, realize opportunities for support and leadership, and remove impediments to implementation.[161]

Accordingly, for consideration by our readers, we include the Report's list of goals and actions:

**Goal 1: Identify a clear pathway toward an adaptable, sustainable, and secure technology marketplace.**

---

[159] Title 1.81.26 (commencing with Section 1798.91.04) to Part 4 of Division 3 of the Civil Code - https://leginfo.legislature.ca.gov/faces/billNavClient.xhtml?bill_id=201720180SB327
[160] *See* Report to the President, *supra* note 107 at 25.
[161] *Id.*

Action 1.1 Using industry-led inclusive processes, establish internationally applicable IoT capability baselines supporting lifecycle security for home and industrial applications founded on voluntary, industry-driven international standards.

Action 1.2 The federal government should leverage industry-developed capability baselines, where appropriate, in establishing capability baselines for IoT devices in U.S. government environments to meet federal security requirements, promote adoption of industry-led baselines, and accelerate international standardization.

Action 1.3 Software development tools and processes to significantly reduce the incidence of security vulnerabilities in commercial-off-the-shelf software must be more widely adopted by industry. The federal government should collaborate with industry to encourage further enhancement and application of these practices and to improve marketplace adoption and accountability.

Action 1.4 Industry should expedite the development and deployment of innovative technologies for prevention and mitigation of distributed threats. Accordingly, where relevant, government should prioritize the application of research and development funds and technology transition efforts to support advancements in DDoS prevention and mitigation, as well as foundational technologies to prevent botnet creation. Where appropriate, civil society should amplify those efforts.

Action 1.5 Government, industry, and civil society should collaborate to ensure that existing best practices, frameworks, and guidelines relevant to IoT, as well as procedures to ensure transparency, are more widely adopted across the digital ecosystem. Emerging risks in the IoT space must be addressed in an open and inclusive fashion.

**Goal 2: Promote innovation in the infrastructure for dynamic adaptation to evolving threats.**

Action 2.1 Internet service providers and their peering partners76 should expand current information sharing to achieve more timely and effective sharing of actionable threat information both domestically and globally.

Action 2.2 Stakeholders and subject matter experts, in consultation with NIST, should lead the development of a CSF Profile for Enterprise DDoS Prevention and Mitigation.

Action 2.3 The federal government should lead by example and demonstrate practicality of technologies, creating market incentives for early adopters.

Action 2.4 Industry, government, and civil society should collaborate with the full range of stakeholders to continue to enhance and standardize information-sharing protocols.

Action 2.5 The federal government should work with U.S. and global infrastructure providers to expand best practices on network traffic management across the ecosystem.

**Goal 3: Promote innovation at the edge of the network to prevent, detect, and mitigate automated, distributed attacks.**

Action 3.1 The networking industry should expand current product development and standardization efforts for effective and secure traffic management in home and enterprise environments.

Action 3.2 Home IT and IoT products should be easy to understand and simple to use securely.

Action 3.3 Enterprises should migrate to network architectures that facilitate detection, disruption, and mitigation of automated, distributed threats. They should also consider how their own networks put others at risk.

Action 3.4 The federal government should investigate how wider IPv6 deployment can alter the economics of both attack and defense.

**Goal 4: Promote and support coalitions between the security, infrastructure, and operational technology communities domestically and around the world.**

Action 4.1 ISPs and large enterprises should increase information sharing with government agencies and with one another to provide more timely and actionable information regarding automated, distributed threats.

Action 4.2 The federal government should promote international adoption of best practices and relevant tools through bilateral and multilateral international engagement.

Action 4.3 Sector-specific regulatory agencies, where relevant, should work with industry to ensure nondeceptive marketing and foster appropriate sector-specific security considerations.

Action 4.4 The community should identify leverage points and take concrete steps to disrupt attacker tools and incentives, including the active sharing and use of reputation data.

Action 4.5 The cybersecurity community should continue to engage with the operational technology community to promote awareness and accelerate incorporation of cybersecurity technologies.

**Goal 5: Increase awareness and education across the ecosystem.**

Action 5.1 The private sector should establish and administer voluntary informational tools for home IoT devices, supported by a scalable and cost-effective assessment process, that consumers can trust and intuitively understand.

Action 5.2 The private sector should establish voluntary labeling schemes for industrial IoT applications, supported by a scalable and cost-effective assessment process, to offer sufficient assurance for critical infrastructure applications of IoT.

Action 5.3 Government should encourage the academic and training sectors to fully integrate secure coding practices into computer science and related programs.

Action 5.4 The academic sector, in collaboration with the National Initiative for Cybersecurity Education, should establish cybersecurity as a fundamental requirement across all engineering disciplines.

Action 5.5 The federal government should establish a public awareness campaign to support recognition and adoption of the home IoT device security baseline and branding.[162]

## X. CONCLUSION

Costly data breaches continue at an alarming rate. The challenge facing humans as they attempt to govern the process of artificial intelligence, machine learning, and the impact of billions of sensory devices connected to the Internet is a challenge to all involved. We believe this Article contributes to our understanding of the widespread exposure to malware associated with the Internet of Things (IoT) and adds to the nascent but emerging literature on governance of enterprise risk, a subject of vital societal importance.

---

[162] *Id.*